\documentclass[aps,floats]{revtex4}
\usepackage{tabularx,graphicx}
\usepackage{epsfig}
\usepackage{amsmath}
\usepackage{amsfonts}
\usepackage{bbm}
\usepackage{graphicx}

\begin{document}

\title{Confining and repulsive potentials from effective non-Abelian gauge 
fields in graphene bilayers}
\author{J. Gonz\'alez  \\}
\address{Instituto de Estructura de la Materia,
        Consejo Superior de Investigaciones Cient\'{\i}ficas, Serrano 123,
        28006 Madrid, Spain}

\date{\today}

\begin{abstract}
We investigate the effect of shear and strain in graphene bilayers, under 
conditions where the distortion of the lattice gives rise to a smooth 
one-dimensional modulation in the stacking sequence of the bilayer. We show
that strain and shear produce characteristic Moir\'e patterns which can 
have the same visual appearance on a large scale, but representing graphene
bilayers with quite different electronic properties. The different features in 
the low-energy electronic bands can be ascribed to the effect of a fictitious 
non-Abelian gauge field mimicking the smooth modulation of the stacking order.
Strained and sheared bilayers show a complementary behavior, which can be 
understood from the fact that the non-Abelian gauge field acts as a repulsive 
interaction in the former, expelling the electron density away from 
the stacking domain walls, while behaving as a confining interaction leading 
to localization of the electronic states in the sheared bilayers. In this 
latter case, the presence of the effective gauge field explains the development 
of almost flat low-energy bands, resembling the form of the zeroth 
Landau level characteristic of a Dirac fermion field. The estimate of the 
gauge field strength in those systems gives a magnitude of the order of several 
tens of Tesla, implying a robust phenomenology that should be susceptible of 
being observed in suitably distorted bilayer samples.

\end{abstract}

\maketitle

\section{Introduction}

During more than a decade, graphene has been attracting much attention due to 
its vast potential for technological application. The one-atom-thick carbon 
layer displays many unconventional electronic properties, that derive to a 
great extent from the peculiar conical dispersion of its valence and conduction 
bands\cite{geim,zhang,neto}. 
The behavior of the electron quasiparticles in graphene is similar to 
that of relativistic massless fermions, which explains the 
appearance of phenomena like the Klein paradox\cite{kats} and the limited 
backscattering experienced by electrons in the material.

A remarkable feature of graphene is also that the interplay between the 
electronic degrees of freedom and the geometry of the lattice makes possible 
to mimic the effect of gauge fields acting on the electron quasiparticles. 
This connection goes back to the description of the fullerene lattices, where 
it has been shown that the degeneracies in the low-energy electronic spectrum 
may be understood from the action of a fictitious non-Abelian gauge field, 
induced by the pentagonal rings in the lattice\cite{np}. 
The consistency of the gauge field construction is certified in that case by 
the correspondence between the degeneracy of the low-lying electronic levels 
and the total flux of the effective magnetic field traversing the surface of 
the molecule.

More recently, the study of the effects of strain on the electron 
quasiparticles has unveiled the possibility of engineering a fictitious gauge
field in the graphene lattice\cite{fic}. 
The distortions of the honeycomb lattice can play a role similar to that of 
a real gauge field, shifting locally the Dirac cones in momentum 
space\cite{mor,man}. 
It has been shown that 
some configurations may actually give rise to an effective magnetic field, 
with a sequence of Landau levels resembling that from a real transverse 
magnetic field\cite{fic}. 
This has been experimentally confirmed when looking at the nanobubbles 
which form in some graphene samples on a substrate, finding signatures of 
effective magnetic fields with magnitudes of the order of $\sim 300$ 
Tesla\cite{crom}.

Yet a number of different effects have been related to the appearance of
fictitious gauge fields in graphene bilayers\cite{mucha,son,prl,mari}. 
In these systems, a 
small amount of strain or shear may give rise to a deviation with respect to 
the perfect registry corresponding to Bernal stacking (so-called $AB$ stacking) 
of conventional bilayer graphene. When the lattice distortion takes the form 
of a smooth one-dimensional modulation of the stacking sequence, the changes 
induced in the low-energy electronic spectrum can be understood as arising 
from an effective non-Abelian gauge field acting on the internal space of two 
Dirac cones (one for each carbon layer)\cite{prl,brey,exper}. 
It has been proposed indeed that some configurations 
with alternating $AB$-$BA$ stacking sequence may lead to largely degenerate 
electronic levels, extending into linear branches of edge states and thus 
providing a clear analogy with the physics of the quantum Hall 
effect\cite{prl}.

The present paper is devoted to discriminate the effect of the fictitious 
non-Abelian gauge fields in different graphene bilayers with smooth domain 
walls between $AB$ and $BA$ stacking. In this respect, we note that a 
sequence of such stacking regions can be formed by applying either shear or 
strain, or in general a combination of both. This leads to stacking sequences 
with Moir\'e patterns which may have the same appearance from a wide 
perspective, but representing graphene bilayers with quite different electronic 
properties. Thus, when the stacking sequence is created by shear, we will see 
that the non-Abelian gauge field manifests as a confining interaction, leading 
to localization of electronic states and the development of almost flat 
low-energy bands. When the stacking sequence is created instead by applying 
strain (pulling in a direction perpendicular to the hexagon rows in the 
honeycomb lattice), we will find that the low-energy electronic states feel 
a repulsive potential arising from the non-Abelian gauge field, showing no 
sign of localization in the band structure.

Our analysis becomes relevant as several experimental observations have 
already shown domain walls between regions of different stacking order in 
graphene bilayers\cite{alden,butz,ju,he}. 
There have been theoretical studies showing that such domain walls may 
support one-dimensional electronic states, once the bilayer electron system 
is gapped in the bulk by applying a transverse electric field\cite{mac,kim}. 
It has been found that such states behave much in the same way as those 
arising at domain walls induced by a change in the interlayer 
electric field\cite{martin,jung}. Regarding our investigation, the 
difference with respect to these studies is that here we address the genuine 
effects of the non-Abelian gauge field characterizing smooth domain walls 
between $AB$ and $BA$ stacking, when no transverse electric field is applied 
to the bilayer. This description may become particularly suitable in those 
cases where the transition between different domains is not abrupt, which 
seems to be common as typical atomic-scale images of the stacking domain 
walls show widths of the order of $\sim 10$ nm.

One of the main findings of our work is that an effective non-Abelian gauge 
field may lead to the development of a zeroth Landau level which is quite 
similar to that arising from the effect of a real transverse magnetic field
on Dirac quasiparticles. The counterpart to this remark is that not all the 
configurations of the non-Abelian gauge field share such a flat band 
characteristic of Landau quantization. This establishes a clear difference 
with respect to the action of an Abelian gauge field, leading to electronic 
features which are nicely illustrated in graphene bilayers with different 
stacking domain walls.

\section{One-dimensional Moir\'e patterns in strained and sheared 
graphene bilayers}

When shear or strain is applied to the sheets of a graphene bilayer, a 
Moir\'e pattern in general appears, which is the reflection of a sequence of 
regions alternating between $AB$ and $BA$ stacking (the conventional stacking 
in bilayer graphene) and $AA$ stacking (where homologous points in the two 
layers fall one on top of each other). We are going to concentrate here on 
patterns where the alternation takes place along two different orthogonal 
directions, which are perpendicular in one case and aligned in the other with 
respect to the hexagon rows in the original layers.

\begin{figure}[h]
\begin{center}
\mbox{\epsfxsize 7.1cm \epsfbox{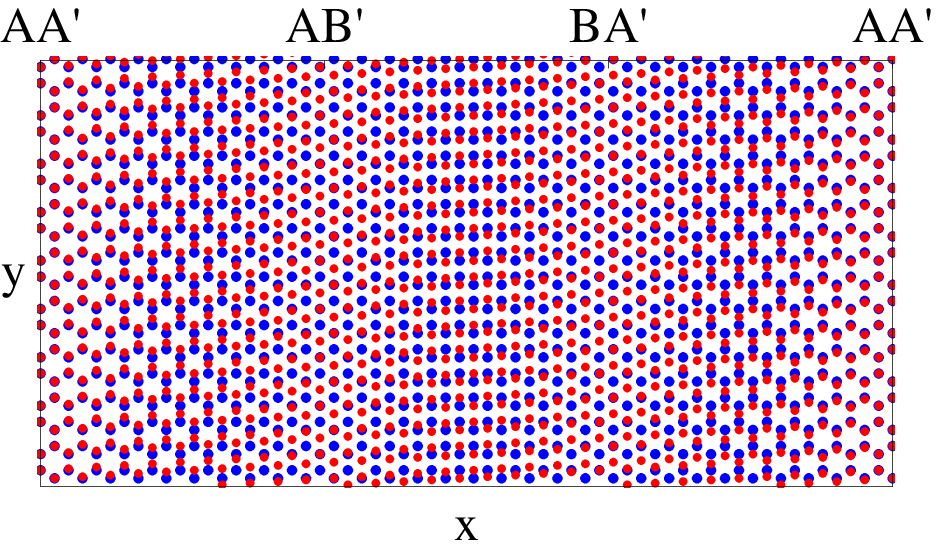}}
\end{center}
\begin{center}
(a)
\end{center}
\begin{center}
\mbox{\epsfxsize 8cm \epsfbox{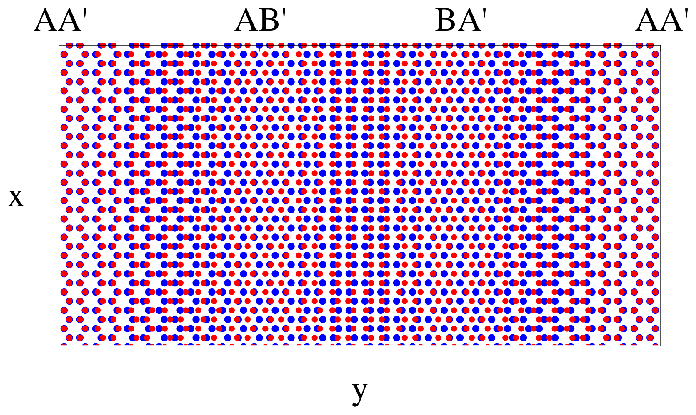}}
\end{center}
\begin{center}
(b)
\end{center}
\caption{Different types of Moir\'e patterns in graphene bilayers, which can 
be obtained (a) by means of shear, pulling laterally the layers in opposite 
directions, and (b) by applying tensile strain, pulling the layers in the 
direction perpendicular to the hexagon rows in the honeycomb lattice.}
\label{one}
\end{figure}

An alternating sequence of $AB$, $BA$ and $AA$ stacking can be obtained by 
applying shear, by means for instance of a lateral shift at the border of 
one of the layers, producing the pattern shown in Fig. \ref{one}(a). But there
is another possibility to produce a sequence of stacking regions by applying 
tensile strain in one of the sheets, in a direction perpendicular to the 
hexagon rows of the honeycomb lattice, leading to the pattern shown in Fig. 
\ref{one}(b). We observe that the Moir\'e patterns shown in Figs. \ref{one}(a) 
and \ref{one}(b) have the same appearance on a large scale, although the 
distortions producing them are quite different at the atomic scale. In general, 
we may think of other Moir\'e patterns with alternating stacking as the result 
of combining strain and shear, leading to one-dimensional periodic sequences 
along directions which do not coincide with the principal axes of the 
original bilayer.

The bilayers shown in Figs. \ref{one}(a) and \ref{one}(b) can be then 
characterized by the different strain supported by the respective lattices. 
This can be quantified in terms of the strain tensors of the two 
layers (labeled by $a = 1, 2$)
\begin{equation}
u^{(a)}_{ij} = \frac{1}{2} (\partial_i u^{(a)}_j + \partial_j u^{(a)}_i 
                         + \partial_i h^{(a)} \partial_j h^{(a)} )
\end{equation}    
where $(u^{(a)}_x, u^{(a)}_y, h^{(a)})$ denotes the three-dimensional 
displacement field of the layer with respect to the equilibrium position. 
Thus, the bilayer shown in Fig. \ref{one}(a) has a strain configuration 
with
\begin{equation}
u^{(a)}_{xy} \neq 0
\label{xy}
\end{equation}
On the other hand, the bilayer in Fig. \ref{one}(b) corresponds to the 
case with
\begin{equation}
u^{(a)}_{yy} \neq 0
\label{yy}
\end{equation}

Strain fields with constant strain tensors given by (\ref{xy}) or 
(\ref{yy}) can already induce slight modifications in the low-energy 
electronic bands of the individual layers. We recall that the effect of strain 
can be mimicked by the action of a gauge field 
on the electron quasiparticles of graphene. The correspondence is such that a 
strain field $u^{(a)}_i$ has the same effect as a vector potential given 
by\cite{man} 
\begin{equation}
\tilde{\mathbf{A}}^{(a)} = \frac{\beta }{a} \left(  \begin{array}{c}
 u^{(a)}_{xx} - u^{(a)}_{yy} \\
   2 u^{(a)}_{xy}  
 \end{array}\right)
\end{equation}
where $a$ stands for the C-C distance and $\beta $ represents the 
variation of the tunneling amplitude with respect to the lattice spacing. 
Then, the bilayer in Fig. \ref{one}(a) is characterized by having an effective 
vector potential with $\tilde{A}^{(a)}_y \neq 0$, while the bilayer in Fig. 
\ref{one}(b) has instead  $\tilde{A}^{(a)}_x \neq 0$. The patterns shown 
in Figs. \ref{one}(a) and \ref{one}(b) can be obtained with just a constant 
$u^{(a)}_{xy} $ or $u^{(a)}_{yy}$. In this case, the strain 
configurations correspond to vanishing pseudomagnetic field, and their 
effect can be seen as a shift of the Dirac cones representing the low-energy 
electronic states.

The important changes in the band structure of the bilayers come however from 
the modulation of the stacking order. In this regard, a remarkable observation 
is that the low-energy bands of the bilayers shown in Figs. \ref{one}(a) and 
\ref{one}(b) have completely different shapes. To illustrate this fact, we can
rely on a tight-binding approximation, taking into account intralayer and 
interlayer tunneling amplitudes to express the hamiltonian of a graphene 
bilayer in the form
\begin{equation}
H_{\rm tb} = - \sum_{i,j} t_{ij} a^\dagger_i a_j 
   - \sum_{i,j} t_{ij} b^\dagger_i b_j
    - \sum_{i,j} \tilde{t}_{ij} a^\dagger_i b_j
\label{tb}
\end{equation}
in terms of electron creation (annihilation) operators $a^\dagger_i (a_i)$ for 
the different sites of the upper layer, and similar operators 
$b^\dagger_i, b_i$ for the lower layer. In (\ref{tb}), one can assume an 
exponential decay to represent the interlayer tunneling amplitudes between sites 
with variable separation\cite{maar}
\begin{equation}
\tilde{t}_{ij}(\mathbf{r}_i-\mathbf{r}_j) = t_0 e^{-|\mathbf{r}_i-\mathbf{r}_j|/a_0}
\label{rule}
\end{equation}
We have chosen in particular $t_0$ and $a_0$ so that the hopping parameter 
between nearest-neighbor sites in each carbon layer is set to 3.2 eV, while 
the hopping parameter between nearest-neighbor sites in different layers (in 
the region of $AB$ stacking) is set to 0.3 eV. In practice, we have reduced the 
complexity of the model by restricting intralayer hopping to nearest-neighbor
sites, and taking a finite range $r_0$ for the interlayer tunneling to allow 
hopping up to a distance equal to the next-to-nearest-neighbor
separation (in the region of $AB$ stacking) between different layers.

\begin{figure}[h]
\begin{center}
\includegraphics[height=3.5cm]{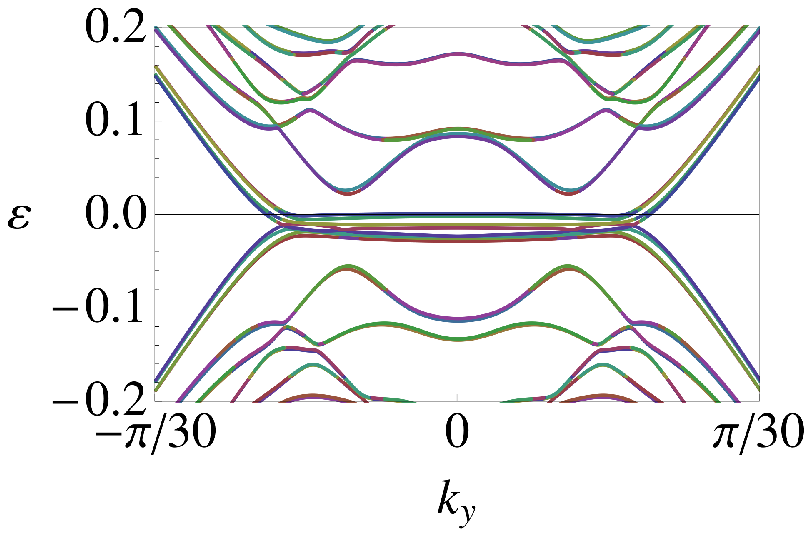}
\hspace{0.5cm}
\includegraphics[height=3.5cm]{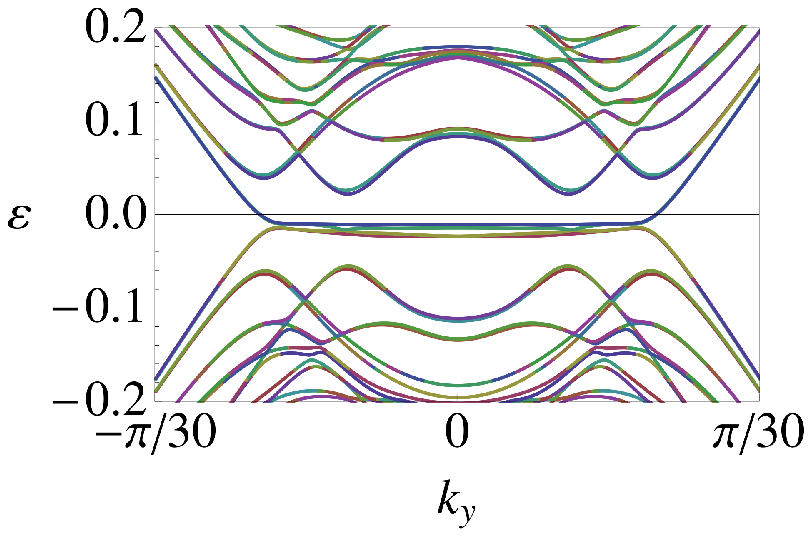}
 \mbox{} \hspace{9.0cm}  (a) \hspace{5.5cm} (b)\\
\mbox{}    \\
\includegraphics[height=3.6cm]{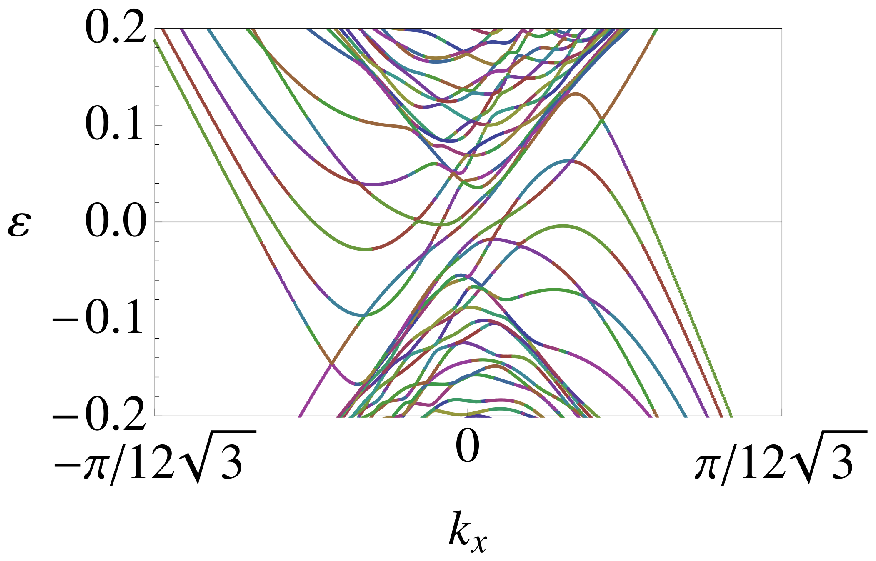}
\hspace{0.3cm}
\includegraphics[height=3.6cm]{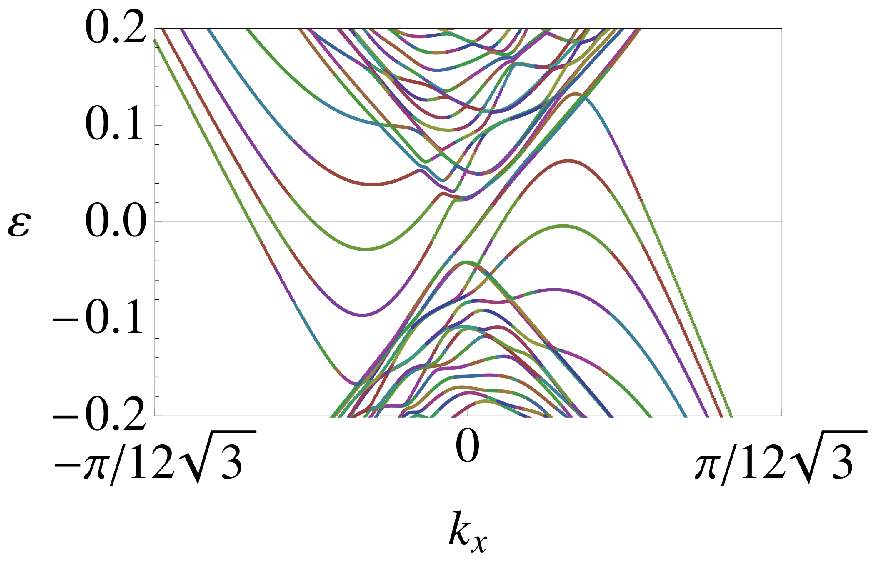}
 \mbox{} \hspace{9.0cm}  (c) \hspace{5.5cm} (d)
\end{center}
\caption{(a),(b): Low-energy bands around the Fermi level of an undoped bilayer 
of the type shown in Fig. \ref{one}(a), obtained by means of a tight-binding 
approximation for a geometry of infinite length along the vertical direction 
and length $L = 210 \sqrt{3}a$ ($a$ being the C-C distance) along the 
horizontal direction, with periodic (a) and open (b) boundary conditions. 
(c),(d): Plots obtained in similar fashion as (a),(b), representing the 
low-energy bands (from a given Dirac valley) of a graphene bilayer of the type 
shown in Fig. \ref{one}(b), for a geometry of infinite length along the 
vertical direction and length $L = 633 a$ along the horizontal direction with 
periodic (c) and open (d) boundary conditions. The energy $\varepsilon $ is 
given in eV and the momentum in units of the inverse of the C-C distance.}
\label{two}
\end{figure}

We can see for instance in Figs. \ref{two}(a) and \ref{two}(b) the low-energy 
bands obtained with the tight-binding approach for a bilayer 
of the type shown in Fig. \ref{one}(a), with infinite length in the $y$ 
direction and length $L = 210 \sqrt{3} a$ ($a$ being the C-C distance) in the 
$x$ direction. The two different plots correspond to taking periodic 
(Fig. \ref{two}(a)) and open (Fig. \ref{two}(b)) boundary conditions.
In the case of a bilayer of the type shown in Fig. \ref{one}(b), 
results from the tight-binding calculation are represented in Figs. 
\ref{two}(c) and \ref{two}(d), for a geometry with infinite length in the $x$ 
direction and length $L = 633 a$ in the $y$ direction. The plots show in this 
case the low-energy bands developing about one of the Dirac valleys, while a 
similar structure (related by mirror symmetry) is to be found at the opposite 
Dirac valley. The two representations differ in the choice of periodic 
(Fig. \ref{two}(c)) and open (Fig. \ref{two}(d)) boundary conditions.

We note that the details of the bands shown in Figs. \ref{two}(a)-\ref{two}(d) 
may slightly depend on the particular range taken for the interlayer 
tunneling. However, it can be checked that the most salient qualitative 
features, like the existence of almost flat bands close to zero energy for the 
sheared bilayer or the parabolic bands for the strained bilayer, 
persist under variations of the parameters used in the tight-binding 
resolution. In this respect, it becomes clear that the effects of strain or 
shear within each layer cannot have a significant impact on the electron 
system at large $L$, since the large period of the Moir\'e pattern implies 
that it must be produced by a small distortion with 
$u^{(a)}_{xy}$ or $u^{(a)}_{yy}$ of order $\sim 1/L$. The great contrast 
between the bands in Figs. \ref{two}(a)-(b) and those in Figs. \ref{two}(c)-(d) 
must be therefore the consequence of some other effect different to that 
associated to the effective gauge field $\tilde{A}^{(a)}_i$. 

As we are going to see, the generic features that appear in the low-energy 
bands of strained and sheared bilayers can be explained from the existence of 
a different type of effective gauge field, which can be recognized in 
the continuum limit approximation to the electron system. This large-$L$ 
approach will allow us to understand why the band structures of the strained 
and the sheared bilayers are so different from each other, despite the similar 
visual appearance for large period of the Moir\'e patterns.

\section{Model of effective non-Abelian gauge fields}

We start by assuming that, in each individual layer (when the interlayer 
coupling is ideally switched off), the low-energy dynamics of electron 
quasiparticles is governed by a Dirac hamiltonian (focusing on a given Dirac 
valley)
\begin{equation}
H^{(a)} =  v_F \boldsymbol{\sigma}\cdot(-i\boldsymbol{\partial}
                                         -  \tilde{\mathbf{A}}^{(a)})
\end{equation}  
with the parameter $v_F$ standing for the Fermi velocity. Moreover, we focus 
on graphene bilayers having Moir\'e patterns with fairly large period, for 
which the coupling between carbon layers can be seen as a smooth spatial 
modulation of the interlayer tunneling. In general, we may discern between 
different interlayer amplitudes depending on the sublattices $A$ and $B$ of
each graphene lattice, and evolving from one stacking domain to the next. 
A simple model can be built by assembling the Dirac 
quasiparticles from the two layers into a four-component spinor 
$\Psi = (\psi^{(1)}_A, \psi^{(1)}_B, \psi^{(2)}_{A'}, \psi^{(2)}_{B'})$. In 
this representation, the hamiltonian accounting for the tunneling 
between layers can be written as 
\begin{equation}
H= v_F \left(\begin{array}{cccc}
0 & -i\nabla^{(1)}_x - \nabla^{(1)}_y  
                           & V_{AA'}(\mathbf{r}) & V_{AB'}(\mathbf{r}) \\
 -i\nabla^{(1)}_x + \nabla^{(1)}_y  & 0 
                              & V_{BA'}(\mathbf{r}) & V_{AA'}(\mathbf{r}) \\
 V_{AA'}^\star(\mathbf{r}) & V_{BA'}^\star(\mathbf{r}) 
                               & 0 &   -i\nabla^{(2)}_x - \nabla^{(2)}_y  \\
 V_{AB'}^\star(\mathbf{r}) & V_{AA'}^\star(\mathbf{r}) 
                                 &  -i\nabla^{(2)}_x + \nabla^{(2)}_y  & 0
   \end{array} \right)
\label{cont}
\end{equation}
where we have introduced the covariant derivatives 
$\nabla^{(a)}_i \equiv \partial_i - i\tilde{A}^{(a)}_i$, and $V_{AA'}, V_{AB'}, 
V_{BA'}$ stand for the interlayer tunneling amplitudes between different 
sublattices.

An important observation is that, assuming that the interlayer potentials are 
real functions, they can be written in terms of gauge fields with off-diagonal 
action on the Dirac quasiparticles of the bilayer\cite{prl}. We 
can introduce the fields $A_x$ and $A_y$ according to the decomposition 
\begin{eqnarray}
V_{AB'}(\mathbf{r})  & = &  - A_x (\mathbf{r}) + A_y (\mathbf{r}) 
                                                   \label{ab}        \\
   V_{BA'}(\mathbf{r})  & = &  - A_x (\mathbf{r}) - A_y (\mathbf{r})   
\label{ba}
\end{eqnarray}
Then we can recast the hamiltonian (\ref{cont}) by thinking of the field
$A_x$ as an off-diagonal shift of the momentum operator $-i \nabla_x$, and 
assigning the same role for $A_y$ with respect to the operator $-i \nabla_y$. 
We can write
\begin{equation}
H = v_F \boldsymbol{\sigma}\cdot(-i\boldsymbol{\nabla}-\hat{\mathbf{A}}) 
             + v_F V_{AA'}\tau_1
\label{rep}
\end{equation}
introducing the vector potential 
\begin{equation}
\hat{\mathbf{A}} =  \left(  \begin{array}{c}
 A_x \tau_1 \\  A_y \tau_2 
 \end{array}\right)
\end{equation}
in terms of a new set of Pauli matrices $\{ \tau_i \}$ acting on the internal 
space of the two layers.

The representation (\ref{rep}) highlights that $\hat{\mathbf{A}}$ is indeed a 
non-Abelian gauge field, as its associated gauge transformations are valued in
the group $SU(2)$. This has also a reflection in the dynamics of the Dirac 
quasiparticles. Disregarding for simplicity the scalar potential $V_{AA'}$ at
this point, we can take the square of the hamiltonian (\ref{rep}) to end up 
with the eigenvalue equation
\begin{equation}
v_F^2 ((-i\boldsymbol{\nabla} - \hat{\mathbf{A}})^2 
             - \sigma_z \hat{F}_{xy} ) \Psi  =  \varepsilon^2  \Psi
\label{eig}
\end{equation}
where the field strength of the non-Abelian gauge potential is\cite{itz}
\begin{equation}
\hat{F}_{ij} = \nabla_{i} \hat{A}_{j} - \nabla_{j} \hat{A}_{i}  
                                -  i [\hat{A}_{i},\hat{A}_{j}]
\label{fij}
\end{equation}
In Eq. (\ref{fij}), $\hat{A}_{i}$ stands for the matrix-valued 
vector potential. The last term of the field strength with the commutator 
provides actually the relevant contribution in the Moir\'e bilayers for large 
period $L$, since the derivatives of the gauge field become then of order 
$\sim 1/L$ and are therefore subdominant in that limit.

\subsection{Confinement from non-Abelian gauge fields in sheared bilayers}

We pay attention first to the case in which shear with constant 
$u^{(a)}_{xy}  \neq 0$ is responsible of the formation of a 
Moir\'e pattern like that in Fig. \ref{one}(a). In the limit of large $L$, we 
may consider the interlayer potentials $V_{AA'}, V_{AB'}$ and $ V_{BA'}$ 
as smooth functions varying only along the $x$ direction. Moreover, we can 
also neglect terms with derivatives of the gauge field $\hat{A}_{i}$, which 
give subdominant contributions of order $\sim 1/L$. To carry out the analysis 
of the effects of the non-Abelian gauge field, we may concentrate on the 
eigenvalue problem (\ref{eig}). Introducing solutions of the form
\begin{equation}
\Psi (\mathbf{r}) = e^{i k_y y } \chi (x)
\end{equation}
we get at large $L$
\begin{equation}
v_F^2 \left( (-i\partial_x - A_x (x) \tau_1 )^2 + (k_y - A_y (x) \tau_2 )^2 
                - 2\sigma_z \tau_3 A_x (x) A_y (x) \right)  \chi (x) = 
                  \varepsilon^2  \chi (x)
\label{eigx}
\end{equation}

Given that the effective gauge field depends only on the $x$ variable, it 
is possible to partially integrate out $A_x $ from (\ref{eigx}) by applying
a gauge transformation $\chi = U \tilde{\chi}$ with
\begin{eqnarray}
U  & = &  \exp \left( i \theta (x) \tau_1  \right)                \\
 \theta (x) & =  &   \int^x ds  A_x (s)
\end{eqnarray}
The eigenvalue equation becomes then
\begin{equation}
v_F^2 \left(-\partial^2_x + U^\dagger (k_y - A_y (x) \tau_2 )^2 U
    - 2\sigma_z U^\dagger \tau_3 A_x (x) A_y (x) U \right)  \tilde{\chi} (x) = 
                  \varepsilon^2  \tilde{\chi} (x)
\label{eigxc}
\end{equation}
Working out the algebra of Pauli matrices, we get from (\ref{eigxc})
\begin{equation}
v_F^2 \left(-\partial^2_x +  (k_y - A_y (x) \hat{n} )^2 
    + (A_y (x) - \sigma_z  A_x (x) \hat{m} )^2  
               - A_x^2 (x) - A_y^2 (x) \right)  \tilde{\chi} (x) = 
                  \varepsilon^2  \tilde{\chi} (x)
\label{eigxsch}
\end{equation}
where we have the matrices with unit square
\begin{eqnarray}
\hat{m} & = & \cos \left( 2 \theta (x) \right) \tau_3  
                  - \sin \left( 2 \theta (x) \right) \tau_2   
                                      \label{csx1}       \\
\hat{n} & = & \sin \left( 2 \theta (x) \right) \tau_3  
                  + \cos \left( 2 \theta (x) \right) \tau_2
\label{csx2}
\end{eqnarray}
The advantage of the expression (\ref{eigxsch}) is that it can be interpreted 
as a Schr\"odinger equation, from which an effective potential 
$V_{\rm eff} (x)$ can be read in terms of the components of the gauge field
\begin{equation}
V_{\rm eff} (x) = (k_y - A_y (x) \hat{n} )^2 
            + (A_y (x) - \sigma_z  A_x (x) \hat{m} )^2  
               - A_x^2 (x) - A_y^2 (x)
\label{veff}
\end{equation}
 
At large $L$, the argument of the cosine and the sine in Eqs. 
(\ref{csx1})-(\ref{csx2}) is
of order $\sim L$ and those functions become very rapidly oscillating. 
We have in any event that $\hat{n}^2 = 1$, so that for vanishing $k_y$
\begin{equation}
\left.  V_{\rm eff} (x) \right|_{k_y = 0}  =   
            (A_y (x) - \sigma_z  A_x (x) \hat{m} )^2 - A_x^2 (x)
\label{veff0}
\end{equation}
The last term in (\ref{veff0}) acts as a confining potential, while the first 
term exerts the opposite effect. Given that $A_x = - (V_{AB'} + V_{BA'})/2$, we 
can anticipate a tendency of the effective gauge field to localize low-energy
states in the regions where the interlayer potentials $V_{AB'}$ and $V_{BA'}$
are not negligible. We have moreover to bear in mind that 
$A_y = (V_{AB'} - V_{BA'})/2$, so that this component becomes small when 
$V_{AB'} \sim V_{BA'}$. This means that, at $k_y = 0$, the low-energy states 
must be preferentially confined in the interface between $AB$ and 
$BA$ stacking.


A similar conclusion can be reached in a more straightforward way when
 $\varepsilon \approx 0$ (at $k_y = 0$). Then we have from the 
hamiltonian (\ref{cont}) (neglecting again for simplicity the scalar potential)
\begin{eqnarray}
-i \partial_x \psi^{(1)}_A + V_{BA'} \psi^{(2)}_{A'}  & \approx &  0         \\
   V_{AB'}^\star  \psi^{(1)}_A - i \partial_x \psi^{(2)}_{A'}  & \approx &  0
\end{eqnarray}
and similar equations for $\psi^{(1)}_B, \psi^{(2)}_{B'}$. At large $L$, we get 
for instance
\begin{equation}
- \partial^2_x \psi^{(1)}_A - V_{BA'} V_{AB'}^\star \psi^{(1)}_A  \approx 0
\end{equation}
which shows that low-energy modes have to be 
confined at the interfaces of the sheared bilayer where $V_{BA'} V_{AB'}^\star$ 
gets larger values.

We recall that the model of effective non-Abelian gauge fields has been already 
used to investigate the behavior of the low-energy bands in sheared bilayers 
with alternating $AA$-$AB$-$BA$ stacking\cite{prl}. In the case of 
an infinite superlattice with such a stacking sequence, it has been shown that 
the model implies indeed the existence of four almost flat bands per Dirac 
valley. We note that this is in agreement with the existence of eight 
approximately flat bands in the low-energy picture shown in Fig. \ref{two}(a), 
taking into account that the plot results from the superposition of the bands 
from the two valleys at $K$ and $K'$ points (which have both momentum 
$k_y = 0$).    

\begin{figure}[h]
\begin{center}
\includegraphics[height=3.5cm]{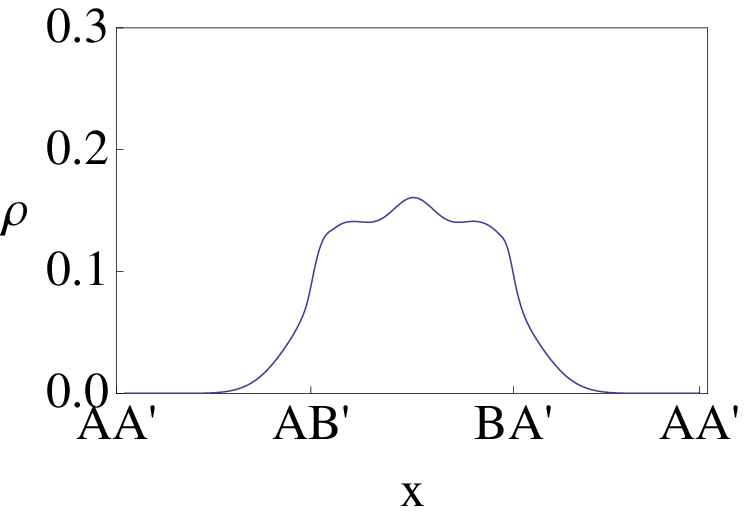}
\hspace{0.5cm}
\includegraphics[height=3.5cm]{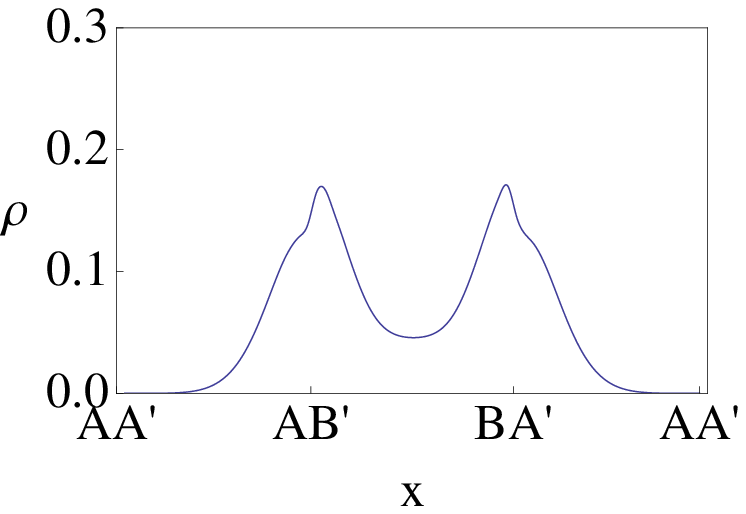}
 \mbox{} \hspace{9.0cm}  (a) \hspace{5.5cm} (b)\\
\includegraphics[height=3.5cm]{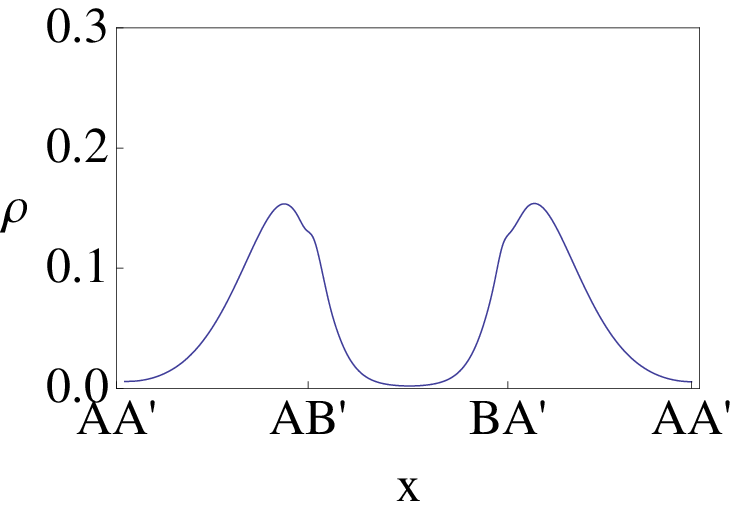}
\hspace{0.5cm}
\includegraphics[height=3.5cm]{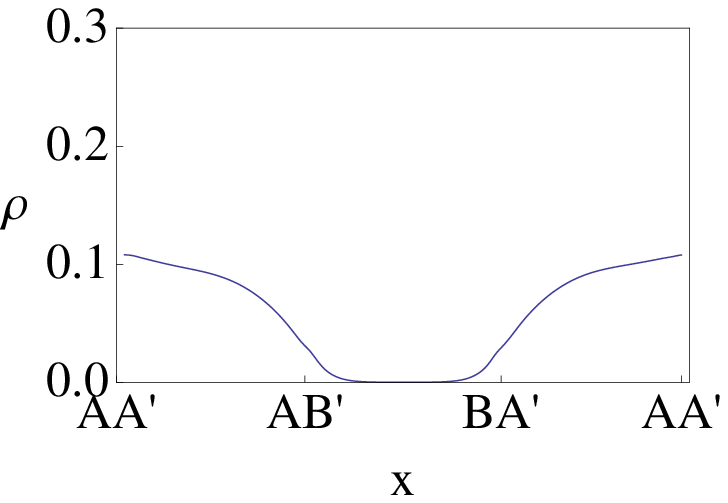}
 \mbox{} \hspace{9.0cm}  (c) \hspace{5.5cm} (d)
\end{center}
\caption{Local density from the states in the four low-energy bands closer to 
$\varepsilon = 0$ in Fig. \ref{two}(b), at respective momenta (from (a)
to (d)) $k_y = 0.01\pi/a, 0.014\pi/a, 0.019\pi/a$ and $0.023\pi/a$ ($a$ being 
the C-C distance).}
\label{three}
\end{figure}

Here we pay attention specifically to the confining properties of 
the gauge field arising from the $AB$-$BA$ domain wall, which are more
properly described in the sheared bilayer with open boundary conditions. In
the plot of Fig. \ref{two}(b), the number of approximately flat bands near
$\varepsilon = 0$ is reduced to four, which is a consequence 
of switching off any confining effect about $AA$ stacking after cutting the 
bilayer in that region. To illustrate the localization of the states, we have 
represented in Fig. \ref{three} a sequence of the local density 
of states from the four approximately flat bands closer to $\varepsilon = 0$ 
in the plot of Fig. \ref{two}(b). We observe that the maxima in the 
local density shift away from the intermediate region connecting $AB$ 
and $BA$ stacking as $k_y$ grows. This is consistent with the expression of 
the potential (\ref{veff}), as it can be seen that the term $- A_y^2$ is 
effectively switched on when the momentum $k_y$ starts deviating from zero. 
Since $A_y = (V_{AB'} - V_{BA'})/2$, that reinforces the confinement already 
induced by the term $- A_x^2$, but moving it towards the regions where either 
$V_{AB'}$ or $V_{BA'}$ have the largest strength.

As observed also from Fig. \ref{three}, the local density of states becomes 
suppressed in the region between $AB$ and $BA$ stacking beyond a certain
momentum $k_y$, when this reaches the dispersive part of the bands with lowest 
energy in Fig. \ref{two}(b). In that regime, the modes in the linear branches 
of such bands can be seen as edge states that are confined to the region where 
the gauge field strength fades away. This draws a consistent 
picture overall, in which the four approximately flat bands in Fig. 
\ref{two}(b) can be interpreted as the zeroth Landau level characteristic of a 
Dirac fermion field, besides other low-energy bands which resemble those 
appearing in graphene lattices placed under a real periodic magnetic 
field\cite{perf}.

\subsection{Repulsion from non-Abelian gauge fields in strained bilayers}

We can adopt a similar large-$L$ approach to explain the low-energy 
properties of the strained bilayers with constant 
$u^{(a)}_{yy}  \neq 0$. In this 
case the interlayer potentials $V_{AA'}, V_{AB'}$ and $ V_{BA'}$ become 
functions only of the $y$ variable, and we can solve the eigenvalue problem
(\ref{eig}) by introducing spinor wavefunctions of the form
\begin{equation}
\Psi (\mathbf{r}) = e^{i k_x x } \chi (y)
\end{equation}
In the limit of large $L$, we get the eigenvalue equation
\begin{equation}
v_F^2 \left( (-i\partial_y - A_y (y) \tau_2 )^2 + (k_x - A_x (y) \tau_1 )^2 
                - 2\sigma_z \tau_3 A_x (y) A_y (y) \right)  \chi (y) = 
                  \varepsilon^2  \chi (y)
\label{eigy}
\end{equation}

We may obtain again an effective Schr\"odinger equation by applying a 
gauge transformation $\chi = U \tilde{\chi }$, with
\begin{eqnarray}
U  & = &  \exp \left( i \theta' (y) \tau_2  \right)                \\
 \theta' (y) & =  &   \int^y ds  A_y (s)
\end{eqnarray}
Eq. (\ref{eigy}) is converted then into
\begin{equation}
v_F^2 \left(-\partial^2_y + U^\dagger (k_x - A_x (y) \tau_1 )^2 U
    - 2\sigma_z U^\dagger \tau_3 A_x (y) A_y (y) U \right)  \tilde{\chi} (y) = 
                  \varepsilon^2  \tilde{\chi} (y)
\label{eigyc}
\end{equation}
Operating with the Pauli matrices, we arrive at
\begin{equation}
v_F^2 \left(-\partial^2_y +  (k_x - A_x (y) \hat{n}' )^2 
    + (A_x (y) - \sigma_z  A_y (y) \hat{m}' )^2  
               - A_x^2 (y) - A_y^2 (y) \right)  \tilde{\chi} (y) = 
                  \varepsilon^2  \tilde{\chi} (y)
\label{eigysch}
\end{equation}
where we have the matrices with unit square
\begin{eqnarray}
\hat{m}' & = & \cos \left( 2 \theta' (y) \right) \tau_3  
                  + \sin \left( 2 \theta' (y) \right) \tau_1     
                                            \label{csy1}         \\
\hat{n}' & = & - \sin \left( 2 \theta' (y) \right) \tau_3  
                  + \cos \left( 2 \theta' (y) \right) \tau_1
\label{csy2}
\end{eqnarray}
The expression (\ref{eigysch}) can be then interpreted as a Schr\"odinger 
equation, providing an effective potential
\begin{equation}
V_{\rm eff}' (y) = (k_x - A_x (y) \hat{n}' )^2 
            + (A_x (y) - \sigma_z  A_y (y) \hat{m}' )^2  
               - A_x^2 (y) - A_y^2 (y)
\label{veffy}
\end{equation}

For very large $L$, the cosine and the sine in $\hat{m}'$ and $\hat{n}'$
oscillate very fast, producing ups and downs that average to zero. We have 
however that $(\hat{n}')^2 = 1$, and we get for vanishing $k_x$
\begin{equation}
\left.  V_{\rm eff}' (y) \right|_{k_x = 0}  =   
       (A_x (y) - \sigma_z  A_y (y) \hat{m}' )^2 - A_y^2 (y)
\label{veffy0}
\end{equation}
The only source of confinement at $k_x = 0$ may come from the term 
$- A_y^2$ in (\ref{veffy0}), but that vanishes for $V_{AB'} = V_{BA'}$. In the 
regions where either $V_{AB'}$ or $V_{BA'}$ have large strength, confinement 
is otherwise compensated by the repulsion exerted by $A_x$ in the first term 
at the right-hand-side of (\ref{veffy0}). It becomes clear then that the 
effective gauge field accounting for the interlayer coupling around $AB$ and 
$BA$ stacking cannot lead generically to confinement of low-energy states in 
the case of the strained bilayer.

Concentrating on the low-energy regime with $\varepsilon \approx 0$ at 
$k_x = 0$, it can be shown more directly that the interlayer potentials 
induce an effect of repulsion at the interface between $AB$ and $BA$ stacking 
in the strained bilayer. From the hamiltonian (\ref{cont}), we get in these 
conditions
\begin{eqnarray}
 \partial_y \psi^{(1)}_A + V_{BA'} \psi^{(2)}_{A'}  & \approx &  0         \\
   V_{AB'}^\star  \psi^{(1)}_A + \partial_y \psi^{(2)}_{A'}  & \approx &  0
\end{eqnarray}
and similar equations for $\psi^{(1)}_B, \psi^{(2)}_{B'}$.
In the limit of large $L$, possible low-energy states should correspond 
therefore to solutions of the equation
\begin{equation}
- \partial^2_y \psi + V_{BA'} V_{AB'}^\star \psi  \approx 0
\label{baab}
\end{equation}
As long as $V_{AB'}$ and $V_{BA'}$ provide a smooth representation of the 
interlayer tunneling, we must have $V_{BA'} V_{AB'}^\star > 0$ in the region
between $AB$ and $BA$ stacking. This means that no low-energy states can be 
bound by the potential in (\ref{baab}), implying that no low-energy states can 
arise from confinement due to the non-Abelian gauge field in the strained 
bilayers.

\begin{figure}[h]
\begin{center}
\includegraphics[height=3.5cm]{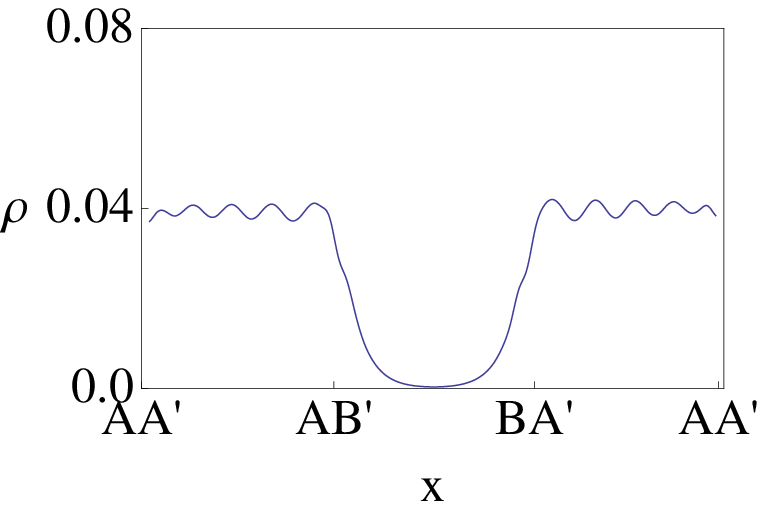}
\hspace{0.5cm}
\includegraphics[height=3.5cm]{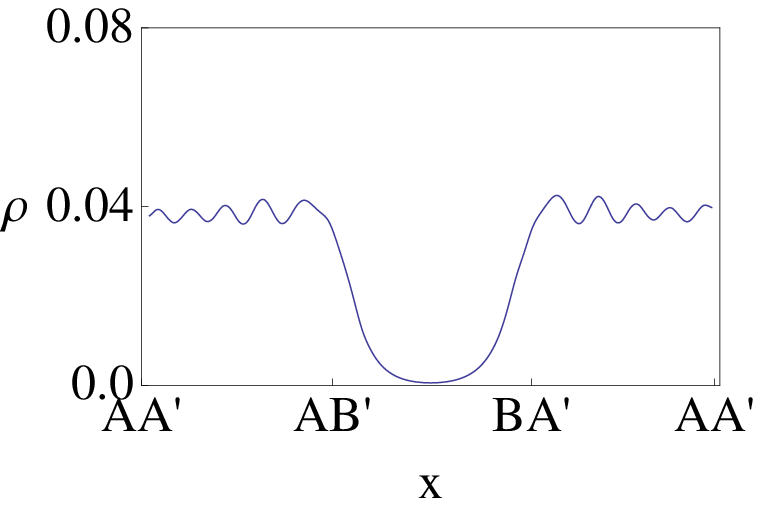}
 \mbox{} \hspace{9.0cm}  (a) \hspace{5.5cm} (b)\\
\mbox{}    \\
\includegraphics[height=3.5cm]{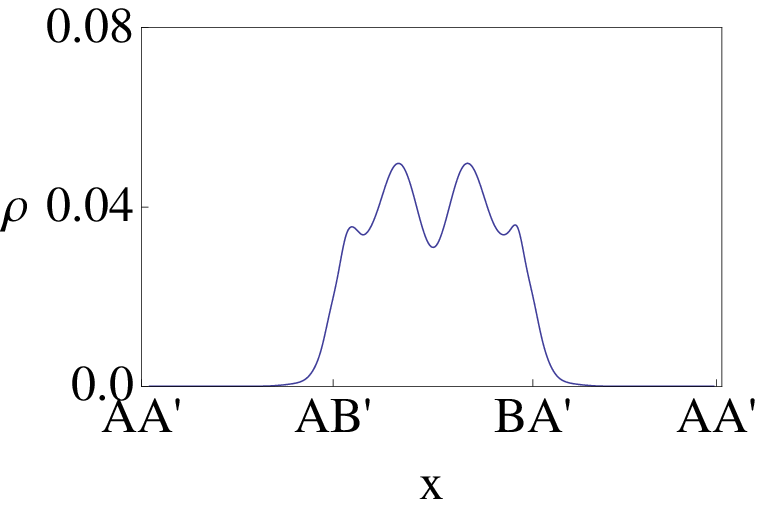}
\hspace{0.5cm}
\includegraphics[height=3.5cm]{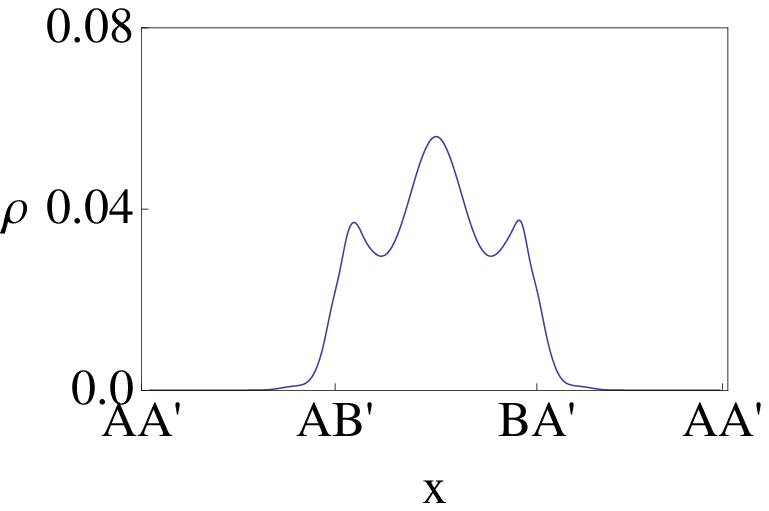}
 \mbox{} \hspace{9.0cm}  (c) \hspace{5.5cm} (d)
\end{center}
\caption{(a),(b): Local density from the states with $k_x = 0$ in the
linear branch close to zero energy (a) and at the bottom of the quadratic
band right above zero energy (b) in the band structure shown in Fig. 
\ref{two}(d). (c),(d): Local density from the states with $k_x = 0.014 \pi/a $ 
($a$ being the C-C distance)
at the top of the quadratic band right below zero energy (c) and at the top of 
the quadratic band right above zero energy (d) appearing to the right in 
the band structure of Fig. \ref{two}(d).}
\label{four}
\end{figure}

The description in terms of the effective gauge field can also account for
the transition to a different regime when we consider momenta 
$k_x \neq 0$. To illustrate this effect, we have represented in Fig. \ref{four}
the local density of states from different bands at vanishing as well as 
nonvanishing momentum $k_x$. As shown in the figure, the local density of 
states with $k_x = 0$ is in general suppressed in the region between $AB$ and 
$BA$ stacking, in accordance with the above arguments. The plots of the local 
density of states with $k_x \neq 0$ (taken from the parabolic bands shown in 
Fig. \ref{two}(d)) display instead a clear confinement in the region mediating 
$AB$ and $BA$ stacking. This is consistent again with the above description, 
since shifting the momentum $k_x$ away from zero amounts to switching on the 
term $- A_x^2$ in the effective potential (\ref{veffy}), leading to attraction 
to the regions where $V_{AB'}$ and $V_{BA'}$ have larger strength.

We observe that the effects of the non-Abelian gauge field in the strained 
bilayer are radically different to those in the sheared bilayer. Looking at 
Figs. \ref{three} and \ref{four}, it may seem that the patterns
are inverted when passing from one type of bilayer to the other. However, it 
has to be stressed that the confinement seen in Figs. \ref{four}(c) and
\ref{four}(d) in the intermediate region between $AB$ and $BA$ stacking 
corresponds to states in parabolic bands whose energy is in general not small. 
This phenomenon is very different to the localization that is displayed by the 
states in the approximately flat bands developed at low energies by the 
sheared bilayers. It is in that case that the effects of the non-Abelian 
gauge field can be assimilated to those of a real Abelian gauge field, 
regarding in particular the formation of flat bands from localized states.

\subsection{Snake states in sheared graphene bilayers}

We want to make contact at this point with the results of previous studies
dealing with stacking domain walls in graphene bilayers. These analyses have
mainly focused on the description of abrupt boundaries between different 
stacking regions, or when there is also a difference in gate voltage 
between the graphene layers\cite{mac,kim}. In those cases, the main finding 
has been that several linear branches appear within the gap in the electronic 
spectrum, connecting the valence and conduction bands of the graphene bilayer. 
It has been shown that these low-energy branches arise as a topological 
effect, which has its origin in the mismatch in the Berry curvature and Chern 
number of the regions connected by the stacking domain wall\cite{mac,kim,jung}. 
In the case of smooth 
modulations of the stacking pattern (with no transverse electric field), we 
are going to see that there are similar low-energy branches, which can be 
understood here as a result of the inversion in the orientation of the gauge 
field strength at the interface between different stacking regions.

\begin{figure}[h]
\begin{center}
\includegraphics[height=3.5cm]{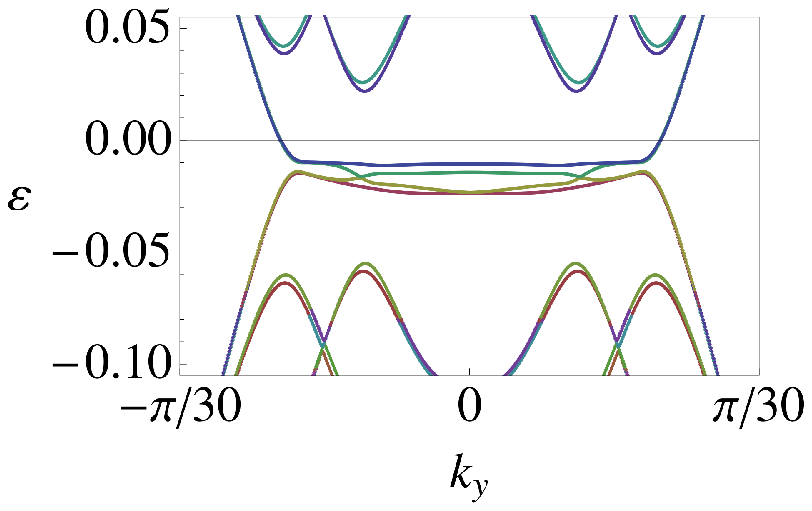}
\hspace{0.5cm}
\includegraphics[height=3.5cm]{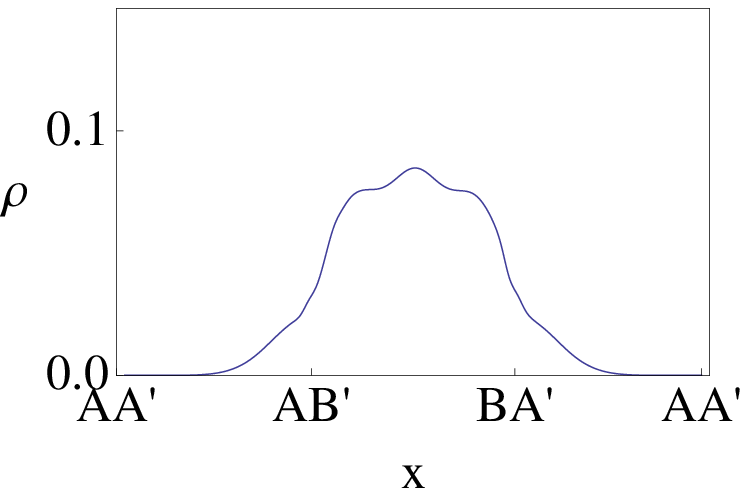}
 \mbox{} \hspace{9.0cm}  (a) \hspace{5.5cm} (b)
\end{center}
\caption{(a) Zoom view of the low-energy part of the band structure shown in 
Fig. \ref{two}(b). The energy $\varepsilon $ is 
given in eV and the momentum in units of the inverse of the C-C distance.
(b) Plot of the local density from the states at the 
crossing of the linear branches within the approximately flat bands in (a).}
\label{five}
\end{figure}

We have already mentioned that, in the limit of large $L$, the field strength 
of the non-Abelian gauge field is dominated by the last term in 
Eq. (\ref{fij}). In the case of the relevant $xy$ component, we get
\begin{equation}
\hat{F}_{xy} \approx  2  A_x (x) A_y (x) \tau_3
\end{equation}
which vanishes at the boundary where 
$V_{AB'} = V_{BA'}$. In cases where one has a modulated magnetic field, the 
lines corresponding to vanishing field strength give rise to effective 
boundaries in the electron system where new edge states may appear. These are 
the so-called snake states, which have been found in a number of situations 
where the spatial modulation leads to an inversion in the orientation of the 
magnetic field\cite{perf,aji,novi}. 
In our graphene bilayers, we have also signatures of snakes 
states, which are already present in the low-energy regime of band structures 
like those shown in Figs. \ref{two}(a) and \ref{two}(b). One has however to 
make a zoom around the approximately flat bands, in order to have an enhanced
view of the relevant features. These are clearly resolved in Fig. 
\ref{five}(a). We observe there the presence of low-energy branches with 
linear crossing, which are the precursor of the linear branches between 
valence and conduction bands already found in the case of gated graphene 
bilayers. The plot of the local density from the states at the crossing point 
between the linear branches, represented in Fig. \ref{five}(b), shows in a 
clear way the confinement of the states in the region between $AB$ and $BA$ 
stacking.

The crossing of the low-energy bands seen in Fig. \ref{five}(a) is indeed a 
direct consequence of the modulated gauge field arising from the $AB$-$BA$ 
stacking pattern. This connection can be established more precisely by 
analyzing the effective gauge field model, adopting a simplified formulation 
in which the scalar potential $V_{AA'}$ is switched off. The spectrum of the 
hamiltonian (\ref{rep}) has particle-hole symmetry, which means that the 
crossing of low-energy linear branches can be characterized from the presence 
of zero-energy modes. Going back to the hamiltonian (\ref{cont}), 
such modes must arise as solutions of the system (assuming that the interlayer
potentials are real)
\begin{equation}
-i \partial_x  \left(\begin{array}{c} \chi^{(1)} \\ \chi^{(2)} \end{array} \right)
= \left(\begin{array}{cc} i k_y  &  - V_{AB'}  \\  - V_{BA'}  &  i k_y  \end{array} \right)
 \left(\begin{array}{c} \chi^{(1)} \\ \chi^{(2)} \end{array} \right)
\label{set}
\end{equation}
Eq. (\ref{set}) can be formally integrated, leading in matrix form to  
\begin{equation}
\left(\begin{array}{c} \chi^{(1)} (x) \\ \chi^{(2)} (x) \end{array} \right) 
= {\rm Pexp} \left\{ i \int_0^x ds \left[ ik_y \mathbbm{1} 
                 + A_x (s) \tau_1 - i A_y (s) \tau_2 \right] \right\}
\left(\begin{array}{c} \chi^{(1)} (0) \\ \chi^{(2)} (0) \end{array} \right)
\label{sol}
\end{equation} 
where ^^ ^^ Pexp" means that the matrix is built from the product of 
exponentials of the differential line elements. While the formal expression 
(\ref{sol}) shows that zero-energy modes may exist, we note that they are 
bound to satisfy a quantization condition, which arises from the boundary 
conditions on the wavefunctions. In the case of bilayers where those 
are imposed for instance at $x=0$ and $x=L$, that amounts to 
enforce the constraint
\begin{equation}
{\rm Pexp} \left\{ i \int_0^L ds \left[ ik_y \mathbbm{1} 
           + A_x (s) \tau_1 - i A_y (s) \tau_2 \right] \right\} = \mathbbm{1}
\label{qc}
\end{equation}
or, less restrictively, the unitarity of the exponential operator in 
(\ref{qc}).

\begin{figure}[h]
\begin{center}
\includegraphics[height=3.5cm]{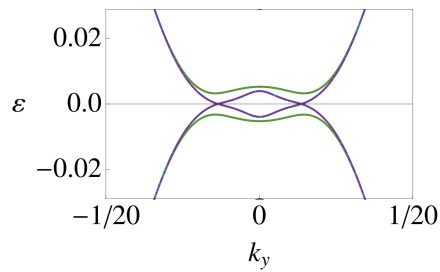}
\end{center}
\caption{Zoom view of the low-energy bands for a sheared bilayer in the 
effective gauge field model, obtained with a single-harmonic approximation to 
the interlayer potentials for $\lambda = 0.1$ eV and 
$\lambda L/v_F = 6.2 \pi $. The energy $\varepsilon $ is 
given in eV and the momentum in units of the inverse of the C-C distance.}
\label{six}
\end{figure}

These considerations can be illustrated most easily in the sheared bilayers 
with periodic boundary conditions. In that case, sensible results can be 
already obtained by taking a single-harmonic approximation for the interlayer 
potentials, which are then represented as 
$V_{AB'}(x) = (\lambda /v_F) [1 + 2 \cos(2\pi x/L - \pi /3)]$ and
$V_{BA'}(x) = (\lambda /v_F) [1 + 2 \cos(2\pi x/L + \pi /3)]$ \cite{prl}. 
By introducing 
these expressions in (\ref{qc}), it can be seen that the boundary condition 
constrains indeed the appearance of the zero modes. For 
$k_y = 0$, one finds for instance that the quantization condition (\ref{qc}) 
is satisfied for values of $L$ such that $\lambda L/v_F = 2\pi n$, with integer 
$n$. When $L$ does not correspond to any ot these values, it is still possible 
to find zero modes for $k_y \neq 0$. This is the instance which is represented 
in Fig. \ref{six}, displaying the low-energy bands in the effective gauge field 
model for $\lambda = 0.1$ eV and $\lambda L/v_F = 6.2 \pi $. The results from 
this simple approximation show to be consistent with the low-energy 
features found in the tight-binding calculation, clarifying the origin of the 
linear crossings in the case of the sheared bilayers.

We remark that snake states appear in the sheared bilayers from the modulation
of the effective gauge field, but also as a consequence of the confining 
character of the interaction that derives from it. This explains that similar
states do not arise in the case of the strained bilayers. As can be seen 
in the band structures shown in Figs. \ref{two}(c) and \ref{two}(d), low-energy
linear branches also exist in those systems. However, their character is very
different to those corresponding to snake states in the sheared bilayers. This
can be already appreciated from the shape of a typical state in the linear 
branches of the strained bilayers shown in Fig. \ref{four}(a), which is a 
reflection of the repulsive interaction expelling the electronic density from 
the region between $AB$ and $BA$ stacking.

\section{Conclusion}

We have seen that the action of strain and shear may lead to graphene 
bilayers which are quite different from the point of view of their low-energy
electronic properties. The mismatch in the registry of the bilayers produces 
characteristic Moir\'e patterns which can look very similar when observed 
from a wide perspective, but may in fact correspond to quite different 
band structures. We have ascribed the different electronic behavior 
to the effect of a fictitious non-Abelian gauge field, which can mimic the 
modulation from $AB$ to $BA$ stacking induced by strain or shear in the 
bilayers.  

We have carried out a comparative analysis of two representative bilayers with
quite different band structures, extending the analyses of previous studies 
about abrupt stacking domain walls in a transverse electric field. We have thus 
seen that strained and sheared bilayers show a complementary behavior, as the 
low-energy states of the former are in general expelled from the region between 
$AB$ and $BA$ stacking, while those of the sheared bilayers tend to be confined 
around that interface. In our effective gauge field model, this
can be understood from the fact that the non-Abelian gauge field may act as a 
repulsive interaction (in the case of the strained bilayers) or as a confining 
interaction leading to localization (in the case of the sheared bilayers).  

The present study becomes pertinent as there have been already 
several experimental observations at the atomic scale of stacking domain walls
in graphene bilayers. These adopt in general the form of smooth modulations in 
the stacking sequence, which can be conveniently described in the framework of 
our continuum approximation. The sample reported in Ref. \cite{he} seems to 
correspond for instance to a sheared bilayer with a sequence of $AB$-$BA$ 
stacking, showing clear signatures of low-energy electronic states around
the stacking domain wall. This is in agreement with the phenomenology 
that we expect from a sheared bilayer. It is quite likely that
more experimental samples of strained or sheared bilayers can be observed in
the future, which may allow to confirm the correspondence we have drawn 
between the atomic arrangement and the low-energy electronic properties of the
bilayers.

From a theoretical point of view, we have elucidated the possibility that a 
non-Abelian gauge field can give rise to a large degeneracy of low-energy 
states, acting much in the same way as a conventional gauge field in the 
quantum Hall regime. The approximately flat low-energy bands we have found in
the sheared bilayers are indeed the analogue of the zeroth Landau level that 
arises in Dirac systems under a strong magnetic field. The lateral
linear branches that are seen at each side in Figs. \ref{two}(a) and 
\ref{two}(b) correspond to edge states, that are here localized at the region 
where the gauge field strength fades away. The magnitude of the effective 
field strength can be easily estimated from the extension of the approximately 
flat bands in Figs. \ref{two}(a)-(b), giving values of the order of 
several tens of Tesla. The experimental signatures of such strong effective 
fields should be then quite robust, leading to a phenomenology susceptible of 
being observed in suitably distorted bilayer samples.

\section*{Aknowledgments}
We acknowledge financial support from MINECO (Spain) through grant No. FIS2014-57432-P.


\begin{thebibliography}{99}



\bibitem{geim}
K. Novoselov, A. Geim, S. Morozov, D. Jiang, M. Katsnelson, I. Grigorieva, 
S. Dubonos, and A. Firsov, Nature {\bf 438}, 197 (2005).

\bibitem{zhang}
Y. Zhang, Y. Tan, H. Stormer, and P. Kim, Nature {\bf 438}, 201 (2005).

\bibitem{neto}
A. H. Castro Neto, F. Guinea, N. M. R. Peres, K. S. Novoselov, and A. K. Geim, 
Rev. Mod. Phys. {\bf 81}, 109 (2009).

\bibitem{kats}
M. I. Katsnelson, K. S. Novoselov, and A. K. Geim, Nature Phys. {\bf 2}, 
620 (2006).

\bibitem{np}
J. Gonz\'alez, F. Guinea, and M. Vozmediano, Nucl. Phys. B {\bf 406}, 
771 (1993).

\bibitem{fic}
F. Guinea, M. Katsnelson, and A. Geim, Nature Phys. {\bf 6}, 30 (2009).

\bibitem{mor}
S.V. Morozov, K.S. Novoselov, M.I. Katsnelson, F. Schedin, L.A. Ponomarenko, 
D. Jiang, and A.K. Geim, Phys. Rev. Lett. {\bf 97}, 016801 (2006).

\bibitem{man}
J. L. Ma\~nes, Phys. Rev. B {\bf 76}, 045430 (2007).

\bibitem{crom}
N. Levy, S. A. Burke, K. L. Meaker, M. Panlasigui, A. Zettl, F. Guinea, 
A. H. Castro Neto, and M. F. Crommie, Science {\bf 329}, 544 (2010).

\bibitem{mucha}
M. Mucha-Kruczy\'nski, I. L. Aleiner, and V. I. Fal'ko, Phys. Rev. B {\bf 84}, 
041404(R) (2011).

\bibitem{son}
Y.-W. Son, S.-M. Choi, Y. P. Hong, S. Woo, and S.-H. Jhi, Phys. Rev. B 
{\bf 84}, 155410 (2011).

\bibitem{prl}
P. San-Jos\'e, J. Gonz\'alez and F. Guinea, Phys. Rev. Lett. {\bf 108}, 
216802 (2012).

\bibitem{mari}
E. Mariani, A. J. Pearce, and F. von Oppen, Phys. Rev. B {\bf 86}, 
165448 (2012).

\bibitem{brey}
For the appearance of effective magnetic fields in one-dimensional graphene
superlattices, see also J. Sun, H. A. Fertig, and L. Brey, Phys. Rev. Lett. 
{\bf 105}, 156801 (2010).

\bibitem{exper}
Experimental evidence for non-Abelian gauge potentials has been also reported 
in twisted graphene bilayers by L.-J. Yin, J.-B. Qiao, W.-J. Zuo, W.-T. Li, 
and L. He, Phys. Rev. B {\bf 92}, 081406(R) (2015).

\bibitem{alden}
J. S. Alden, A. W. Tsen, P. Y. Huang, R. Hovden, L. Brown, J. Park, 
D. A. Muller, and P. L. McEuen, Proc. Natl. Acad. Sci. USA {\bf 110}, 
11256 (2013).

\bibitem{butz}
B. Butz, C. Dolle, F. Niekiel, K. Weber, D. Waldmann, H. B. Weber, B. Meyer, 
and E. Spiecker, Nature {\bf 505}, 533 (2014).

\bibitem{ju}
L. Ju, Z. Shi, N. Nair, Y. Lv, C. Jin, J. Velasco Jr, C. Ojeda-Aristizabal, 
H. A. Bechtel, M. C. Martin, A. Zettl, J. Analytis, and F. Wang, 
Nature {\bf 520}, 650 (2015).

\bibitem{he}
L.-J. Yin, H. Jiang, J.-B. Qiao, and L. He, Nature Commun. {\bf 7}, 11760 
(2016).

\bibitem{mac}
F. Zhang, A. H. MacDonald, and E. J. Mele, Proc. Natl. Acad. Sci. USA 
{\bf 110}, 10546 (2013).

\bibitem{kim}
A. Vaezi, Y. Liang, D. H. Ngai, L. Yang, and E.-A. Kim, 
Phys. Rev. X {\bf 3}, 021018 (2013).

\bibitem{martin}
I. Martin, Ya. M. Blanter, and A. F. Morpurgo, Phys. Rev. Lett. {\bf 100},
036804 (2008).

\bibitem{jung}
J. Jung, F. Zhang, Z. Qiao, and A. H. MacDonald, Phys. Rev. B {\bf 84}, 
075418 (2011).


\bibitem{maar}
A. A. Maarouf, C. L. Kane, and E. J. Mele, Phys. Rev. B {\bf 61}, 
11156 (2000).

\bibitem{itz}
C. Itzykson and J. B. Zuber, Quantum Field Theory (McGraw Hill, 
New York, 1985), Chap. 12.

\bibitem{perf}
E. Perfetto, J. Gonz\'alez, F. Guinea, S. Bellucci, and P. Onorato,
Phys. Rev. B {\bf 76}, 125430 (2007).

\bibitem{aji}
H. Ajiki and T. Ando, J. Phys. Soc. Jpn. {\bf 62}, 1255 (1993); 
{\bf 65}, 505 (1996).

\bibitem{novi}
H.-W. Lee and D. S. Novikov, Phys. Rev. B {\bf 68}, 155402 (2003).





\end{thebibliography}
\end{document}